\documentclass[twocolumn, trackchanges]{aastex631}

\usepackage{amsmath}
\usepackage{multirow}

\newcolumntype{L}[1]{>{\raggedright\let\newline\\\arraybackslash\hspace{0pt}}m{#1}}
\newcolumntype{C}[1]{>{\centering\let\newline\\\arraybackslash\hspace{0pt}}m{#1}}
\newcolumntype{R}[1]{>{\raggedleft\let\newline\\\arraybackslash\hspace{0pt}}m{#1}}

\received{Jan , 2021}
\revised{Jan, 2021}
\accepted{\today}
\submitjournal{ApJ}

\shorttitle{Mass-function of binary black hole systems }
\shortauthors{Park $\&$ Kim $\&$ Kim et al.}

\graphicspath{{./}{figures/}}

\begin{document}

\title{On the mass-function of GWTC-2 binary black hole systems and their progenitors}
\correspondingauthor{Maurice Van Putten}

\author{Hye-Jin Park}
\altaffiliation{These authors contributed equally to this work.}
\affiliation{Research School of Astronomy and Astrophysics, Australian National University, Cotter Road, Weston Creek, ACT 2611, Australia}

\author{Shin-Jeong Kim}
\altaffiliation{These authors contributed equally to this work.}
\author{Shinna Kim}
\altaffiliation{These authors contributed equally to this work.}
\author{Maurice H.P.M. van Putten}
\thanks{mvp@sejong.ac.kr}
\affiliation{
Department of Physics and Astronomy, Sejong University, 209 Neungdong-ro, Gwangjin-gu, Seoul 05006, Republic of Korea
}

\begin{abstract}
The distribution of LIGO black hole binaries (BBH) shows an intermediate-mass range consistent with the Salpeter Initial Mass Function (IMF) in black hole formation by core-collapse supernovae, subject to preserving binary association. They are effectively parameterized by mean mass $\mu$ with Pearson correlation coefficient $r = 0.93\,\pm\,0.06$ of secondary to primary masses with mean mass-ratio $\bar{q}\simeq 0.67$, $q=M_2/M_1$, consistent with the paucity of intermediate-mass X-ray binaries. 
The mass-function of LIGO BBHs is well-approximated by a broken power-law with a tail $\mu\gtrsim 31.4M_\odot$ {in the mean binary mass $\mu=\left(M_1+M_2\right)/2$}.
 Its power-law index \textbf{$\alpha_{B,true}=4.77\pm 0.73$} 
 inferred from the tail of the observed mass-function
 is found to approach the upper bound $2\alpha_S=4.7$ of the uncorrelated binary initial mass-function, defined by the Salpeter index $\alpha_S=2.35$ of the Initial Mass Function of stars. The observed low scatter in BBH mass ratio $q$ evidences equalizing mass-transfer in binary evolution prior to BBH formation. At the progenitor redshift $z^\prime$, furthermore, the power-law index satisfies $\alpha_B^\prime>\alpha_B$ in a flat $\Lambda$CDM background cosmology.  
 The bound $\alpha_{B,true}^\prime \lesssim 2\alpha_S$ hereby precludes early formation at arbitrarily high redshift $z^\prime \gg1$, that may be made more precise and robust with extended BBH surveys from upcoming LIGO O4-5 observations. 
\end{abstract}


\section{Introduction}\label{sec:intro}

It is well known that stellar black holes (BHs) are formed by gravitational collapse or fallback from core-collapse supernovae of high-mass stars (\citealt{1939PhRv...56..455O}; \citealt{1966PhRv..141.1232M,1990RvMP...62..801B,1999ApJ...522..413F,fryer2001theoretical,2012ARNPS..62..407J}).
A key open question is their mass-function and its relation to the Salpeter stellar initial mass function (IMF) 
(\citealt{salpeter1955luminosity}),
\begin{equation}
\label{eq:imf}
dN \propto M^{-\alpha_S}\,dM,
\end{equation}
where $M$ is the mass in units of $M_{\odot}$, $dN$ is the number of stars in the corresponding mass range and $\alpha_{S}$ = 2.35 (\citealt{salpeter1955luminosity}). In the case of $M$\,$>$\,5 $M_{\odot}$, Eq.~(\ref{eq:imf}) appears to be remarkably universal and robust against variations in metallicity following studies of star forming regions of the Large Magellanic Cloud with Z\,$=$\,0.008 (\citealt{dario2009complete,gouliermis2006low}), though some dependencies may exist more generally (\citealt{2012ApJ...760...71C,2013ApJ...771...29G,2013ApJ...763..110K}). 

\begin{figure} 
\centering
\includegraphics[width = 0.9\linewidth]{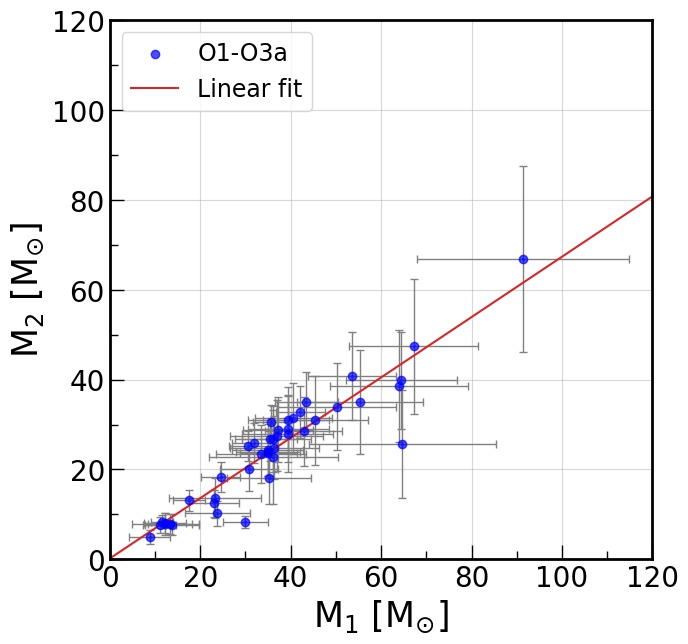}
\caption{Overview of the primary $(M_1)$ and secondary $(M_2)$ mass estimates of the intermediate BBHs in GWTC-2. 
Grey crosses indicate uncertainties. A linear fit shows a slope 0.67 with a correlation coefficient $r=0.93\pm0.06$ (\S\ref{IBMF}).}
\label{fig:catalog}
\end{figure}

Through a complex process of massive stellar binary evolution, Eq.~(\ref{eq:imf}) is expected to have an imprint on the mass-function of their offspring in binary black holes (BBHs) recently observed by LIGO\footnote{The Laser Interferometer Gravitational-Wave Observatory}-Virgo (Fig.~\ref{fig:catalog}). Starting from Eq.~(\ref{eq:imf}) normalized to $\xi\,=\,M/M_{\odot}$, the cumulative count $\propto M^{-\alpha +1}$ and mass $\propto M^{-\alpha +2}$ depends on the lower cut-off $\xi_{c}\,=\,M_{c}/M_{\odot}$, where $M_{c}$ denotes the minimum for  hydrogen ignition. Slightly above the Jupiter mass, $0.01<\xi_{c}<0.1$, the precise value of which is uncertain. 
We assume $\xi \sim 20$ for the low mass cut-off to produce BHs \citep[e.g.][]{1999ApJ...522..413F}, whereby
\begin{equation}
\begin{array}{l}
{M_{\rm{BH}}}/{M_{\rm{tot}}} = (\xi_c/\xi)^{1.35} \sim 7-16\,\%, \\  
{N_{\rm{BH}}}/{N_{\rm{tot}}} = (\xi_c/\xi)^{0.35} \sim 0.004-0.08\,\%.
\end{array}
\end{equation}
Observational support for these percentages is found in population studies of the Milky Way \citep{elbert2018counting}, showing a number fraction of BHs of about 0.09\,$\%$ with aforementioned $\xi_{c}\simeq0.1$. This implies an expected mean BH-progenitor stellar mass
$\bar{M}_{\rm{BH}} = M_{\rm{BH}}/N_{\rm{BH}}$ = $\left(M_{\rm{BH}}/M_{\rm{tot}}\right)$ $\left(N_{\rm{BH}}/N_{\rm{tot}}\right)^{-1}$ $\bar{M}_{*}$,
where $\bar{M}_{*} = 0.1\--0.5\,M_{\odot}$ is the mean stellar mass. 
For the Milky Way,  $\bar{M}_{*}/N_{\rm{tot}}$ can be estimated from the stellar population by total mass $5\times10^{10}M_{\odot}$ \citep{licquia2015improved} and count $(1-5)\times10^{11}$. Based on $\bar{M}_{\rm{BH}}$, the typical BH-progenitor masses are somewhere between 20 and $900\,M_{\odot}$, that we summarize by the geometric mean
\begin{equation}
{M}_{0}\!\,=\sqrt{20\times900} \approx 137\,M_{\odot}.
\label{EQN_geo}
\end{equation}

As a progenitor stellar mass, Eq.~(\ref{EQN_geo}) is subject to mass-loss in the evolution of their progenitor binaries of massive stars.
Preserving binary association limits mass-loss, in winds from the upper atmosphere of a star and in the core-collapse supernova producing the BHs. Regarding the first, simulations predict a positive correlation of mass loss with stellar mass, which tends to suppress the population of heavier stars \citep[ameliorated by their relatively shorter lifetimes,][]{2012ApJ...757...91B} and restrict the maximum BH mass produced in the final core-collapse supernova \citep{fryer2001theoretical}. To illustrate the second, consider two (consecutive) spherically symmetric explosions in circular binaries, neglecting kick velocities and induced ellipticity \citep{1975ApJ...200..145W}. Mass-loss in each is bounded by one-half the total mass for the binary to survive. Let $0<\epsilon<1$ denote the fractional mass loss in each supernova. As the more massive star explodes ($M_2^\prime \le M_1^\prime$), it blows off a stellar mass fraction $\epsilon$ leaving a BH mass $M_{1} \le (1-\epsilon)\,M'_{1}$ and total mass of the binary
$\frac{1}{2}\left( M_1^\prime + M_2^\prime\right)< M''\le M'_{2} + (1-\epsilon)\,M'_{1}$, where the single prime now refers to masses just prior to the supernova event, after mass-loss in winds.
The condition for the binary to survive implies $\epsilon<(1+M_2^\prime/M_1^\prime)$. 
The explosion of the secondary similarly must satisfy
$\epsilon\,M'_{2} < \frac{1}{2}M^{\prime\prime} \leq \frac{1}{2} (M'_{2} + (1-\epsilon)\,M'_{1})$,
giving the more restrictive condition 
$\epsilon < \min\left\{\frac{2}{3},\frac{1}{2}\left(1+{M_2^\prime}/{M_1^\prime}\right)\right\}$.
In the approximation that the mass-loss fraction $\epsilon$ in the progenitor stars suppresses the mass of their BH remnants, 
approximately equal-mass stellar binaries are expected to produce 
\begin{eqnarray}
\bar{M}\simeq \left(1-\epsilon\right)M_0 
\simeq 46\,M_\odot.
\label{EQN_geo1}
\end{eqnarray}
A very similar result derives from a putative stellar progenitor binary of, say, 100\,$M_{\odot}$ and 200\,$M_{\odot}$ with geometric mean Eq.~(\ref{EQN_geo}). Mass-loss by the supernova explosion at the limit $\epsilon=2/3$ produces a binary with intermediate masses $M_1=33\,M_{\odot}$ and $M_2=67\,M_{\odot}$ with geometric mean $47\,M_\odot$ - illustrative for the intermediate BBH range in GWTC-2 (Fig.~\ref{fig:catalog}). 

To date, intermediate-mass black holes (IMBHs) such as Eq.~(\ref{EQN_geo1}) have remained elusive in electromagnetic (EM) observations for reasons not well understood. However, such may derive from a mass-correlation in binary stellar progenitors.
If so, BBH systems form rather promptly with high-mass BHs with high-mass stellar companions practically undetectable (in the electromagnetic spectrum) by the short lifetime of the latter. Indeed, the observed black hole masses of X-ray binary systems are $(3-18)\times M_{\odot}$ (e.g., \citealt{remillard2006x}; \citealt{miller2014astrometric}) - \emph{much} smaller than the geometric mean in Eq.~(\ref{EQN_geo}) predicted by the Salpeter IMF. 

Over the past few years, however, a new window opens to probe BBH systems by gravitational radiation
(Fig.~\ref{fig:catalog}).
The GW150914 event, the first LIGO-detected, is a merger of the BBH system with masses of 36\,$M_\odot$ and 29\,$M_\odot$ (\citealt{abbott2016observation}). The second Gravitational-Wave Transient Catalog (GWTC-2; \citealt{abbott2020gwtc}) of the LIGO-Virgo collaboration provides a unique opportunity for us to study this high-mass population of BBHs. The most massive BH merger event, GW190521, comprises 91.4\,$M_{\odot}$ and 66.8\,$M_{\odot}$ - \emph{not} surprising given Eq.~(\ref{EQN_geo}).
 
A key question is the cosmological origin in binary stellar progenitors of the BBHs, notably 
relative to the peak in the cosmic star formation rate around redshift $z\simeq1.9$ (\citealt{2014ARAA..52..415M}).
To address whether the majority of the progenitor systems are found before or after, we study the binary mass-function of BHs of GWTC-2 in the context of Eq.~(\ref{eq:imf}) and possible effects of cosmic time-dilation.

To start, we consider some statistical properties and mass-function of the intermediate-mass BBHs in GWTC-2 (\S\ref{IBMF}), focused on power-law behavior in the tail of this population. Cosmic time-dilation effects on the power-law index is studied in \S\ref{sec:cosm}. We summarize our results in \S\ref{sec:summary}.

\section{Mass-function index in GWTC-2}\label{IBMF}

We consider LIGO compact binary mergers excluding events involving neutron stars (GW170817, GW190425 and GW190814) based on estimated masses, and three more by limited False Alarm Rates (\citealt{abbott2019gwtc}; \citealt{abbott2020gwtc}). 

In the BBHs in GWTC-2 (Fig.~\ref{fig:catalog}), the primary ($M_{1}$) and secondary masses ($M_{2}$) are strongly correlated with Pearson correlation coefficient $r$ and slope $s$,
\begin{equation}
r = 0.93\,\pm\,0.06,~~s=0.67,
\label{EQN_r}
\end{equation} 
representing a mean $\bar{q}$ of their mass-ratio $M_2/M_1$. 
The minimum and maximum of the mean binary masses are 6.9\,$M_{\odot}$ and 79.1\,$M_{\odot}$, respectively. 

\begin{figure} 
\centering
\includegraphics[width=0.8\columnwidth]{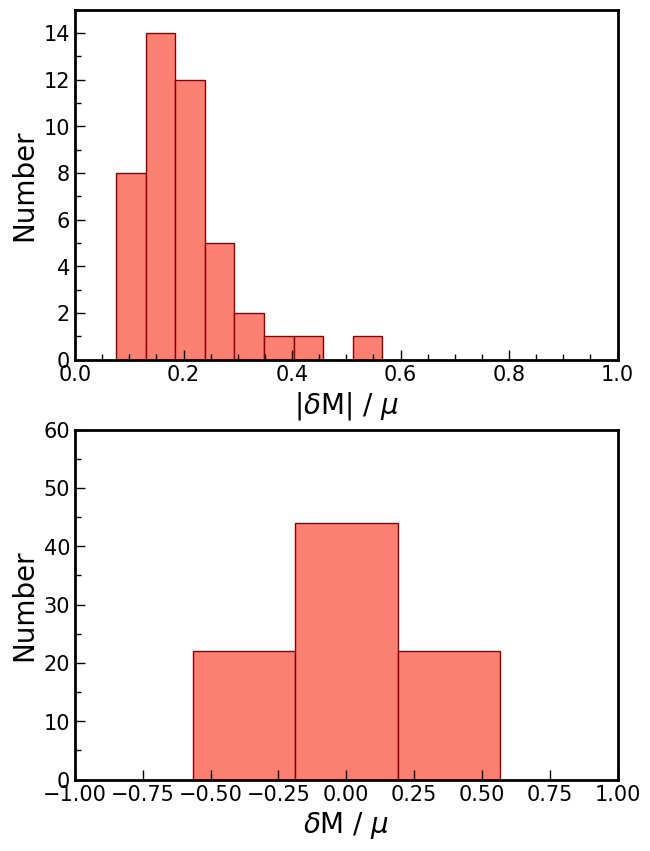}
\caption{(Top panel.) Histogram of normalized mass differences in the BBHs of GWTC-2, $\left| \delta M/\mu\right|$. (Bottom panel.) Symmetric histogram of signed normalized mass difference (over all individual BH masses) in three bins with standard deviation 22\%.}
\label{fig:norm_delm}
\end{figure}

Fig.~\ref{fig:norm_delm} shows the normalized mass difference $\delta M/\mu$, $\delta M = M_1-\mu$, {where 
\begin{eqnarray}
\mu=\frac{1}{2}\left(M_1+M_2\right)
\label{EQN_mudef}
\end{eqnarray}
refers to the mean binary mass.} Shown is a distinctly narrow skewed Gaussian mass-distribution, evidencing a relatively small mass-difference between primary and secondary in these binaries. The relatively modest standard deviation of 22\% in the signed mass-differences (Fig.~\ref{fig:norm_delm}) shows the BBHs are composed of roughly equal-mass BHs, leaving the mean mass as the primary parameter in their mass-distribution.

Figs. \ref{fig:catalog}-\ref{fig:norm_delm} highlight a rather tight correction between the two masses, see also \cite{2020ApJ...891L..27F}. This permits parameterizing BBHs by their mean mass $\mu$. After all,
their observation is determined by their luminosity in gravitational-wave emission, primarily as a function of chirp mass, satisfying
\begin{eqnarray}
{\cal M} = \frac{M_1^{3/5}M_2^{3/5}}{\left(M_1+M_2\right)^{1/5}}\simeq 2^{-{1}/{5}}\mu\left(1-\frac{3}{5}x^2\right),
\label{EQN_Mc}
\end{eqnarray} 
where $x=\delta M/\mu$. The distribution of $\delta M/\mu$ shown in Fig.~\ref{fig:norm_delm} shows $x^2\simeq 0.04$. 
By (\ref{EQN_Mc}), the chirp mass ${\cal M}$ of our BBH binaries in Fig.~\ref{fig:catalog} 
is tracked by the mean mass $\mu$ to better than 10\%. In light of this and aforementioned slope in Fig.~\ref{fig:catalog}, 
$\mu$ provides a statistically equivalent parameterization to primary mass $M_1$ \citep{2021ApJ...913L...7A}, satisfying
\begin{eqnarray}
\mu = \kappa\, M_1, 
\label{EQN_kappa}
\end{eqnarray}
where $\kappa = \frac{1}{2}\left( s + 1\right)=0.8350$, ignoring any potential mass-dependence in this correlation.

We next turn to the observed and true mass-functions of the BBH mergers \citep{2021ApJ...913L...7A}.

\section{Broken power law mass distribution}

\begin{figure*} 
    \centering
    \includegraphics[width =0.89\textwidth]{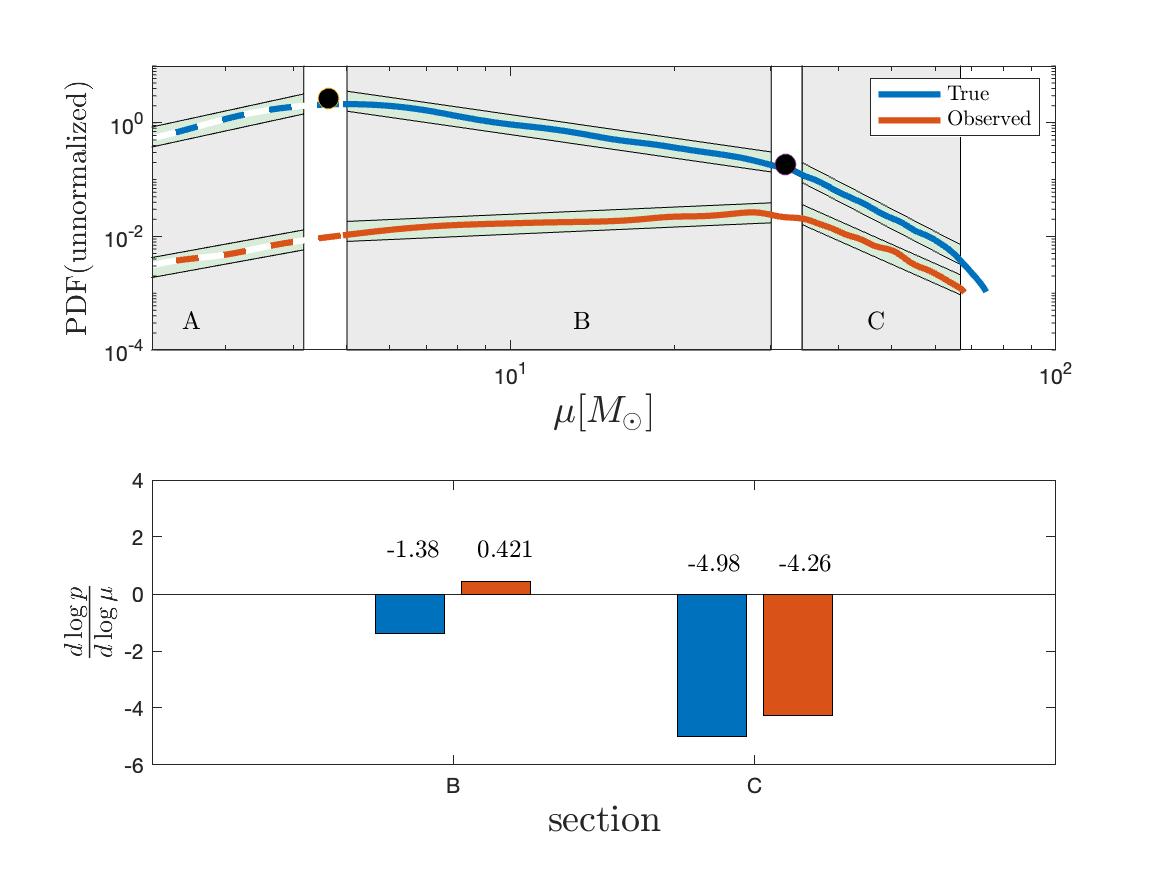}
    \caption{(Top panel.) The true and observed BBH mass-functions of GWTC-2 modeled by a broken power distribution \citep{2021ApJ...913L...7A}, shown in a loglog plot rescaled to $\mu$ according to 
    (\ref{EQN_kappa}) in units of Gpc$^{-3}$yr$^{-1}M_\odot^{-1}$ and, respectively, $M_\odot^{-1}$.
    Breaks shown are at $\mu_{b1}=4.64M_\odot$ and $\mu_{b2}=31.4M_\odot$ (solid black dots). (By \ref{EQN_kappa}, equivalent values $M_1=5.56M_\odot$ and $M_1=37.6M_\odot$ obtain, the latter consistent with $m_{break}=39.7_{-9.1}^{20.3}M_\odot$ in \cite{2021ApJ...913L...7A}.)
    Section A refers to a model extension, below the observed BBH mass-range $6.9M_\odot \le \mu \le 79.1M_\odot$.
    (Lower panel.) The tail of the distribution corresponds (Section C) shows a power-law index 
    $-4.98$ of the true distribution steeper by about 17\% than 
    $-4.26$ of the observed distribution.}
\label{fig:plaw}
\end{figure*}

\begin{figure*} 
    \centering
    \includegraphics[width =0.49\textwidth]{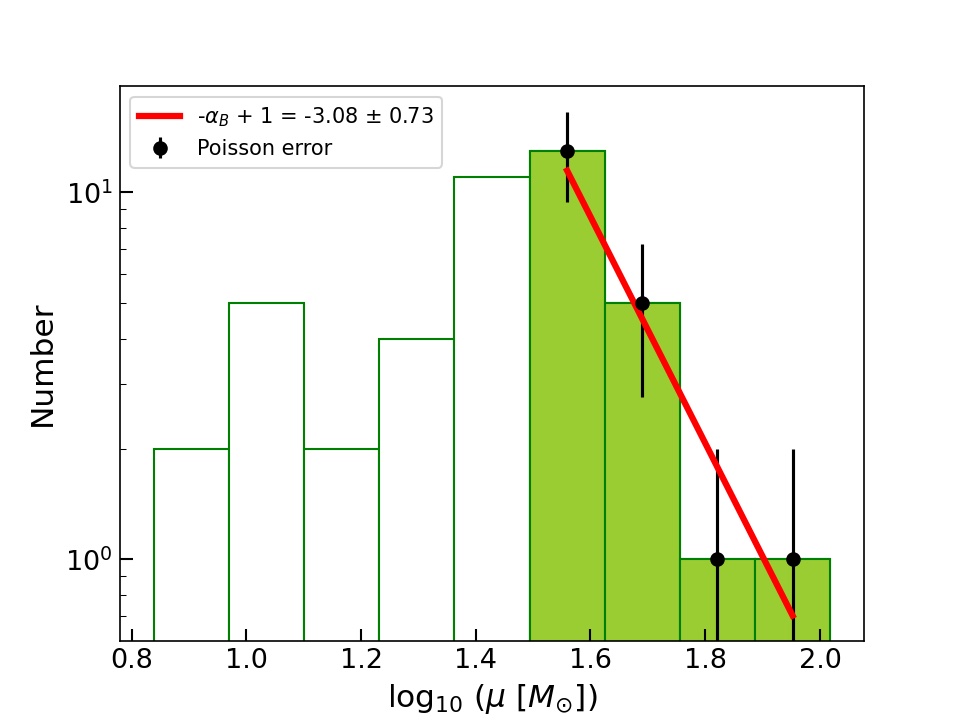}
    \includegraphics[width =0.49\textwidth]{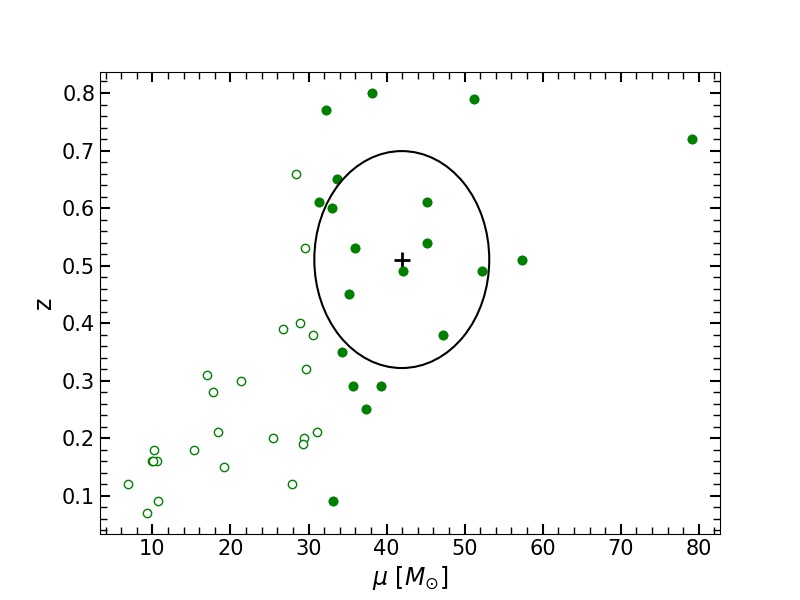}
    \caption{(Left panel.) The binned PDF of the BBH mass-function by mean mass $\mu$ of BBH systems in GWTC-2. The tail of the $\mu$-distribution beyond a peak at $\gtrsim\,31.4\,M_\odot$ (solid green) is fitted to derive a power-law index $\alpha_B$ indicated by the straight line (red with black uncertainties by counting statistics) in the bins with 13, 5, 1, and 1 element(s). (Right panel.) $\mu$-redshift relation (circles) with elements in the tail highlighted (filled). Mean values $(\bar{\mu},\bar{z})=(41.9\,M_\odot,0.51)$ of mass and redshift in the tail (+) is indicated with $1\sigma$ uncertainty (ellipse). }
\label{fig:hist}
\end{figure*}

LIGO BBHs mergers show a power law behavior in their true (astrophysical, per unit volume and observation time) mass distribution inferred from the observed distribution, here shown in Fig.~\ref{fig:plaw} - a loglog plot of data from Figs. 3-4 of \cite{2021ApJ...913L...7A}.

Here, our focus is on the tail $\mu \geq \mu_{b2}=31.4M_\odot$, shown in Section C of Fig.\,
\ref{fig:plaw}, rather than the intermediate mass range of Section B ($\mu_{b1}=4.64M_\odot < \mu < \mu_{b2}$). The location of the second break is consistent with $\mu_{break}=33.15M_\odot$ estimated by (\ref{EQN_kappa}) from $m_{break}=39.7_{-9.1}^{20.3}M_\odot$ in \cite{2021ApJ...913L...7A} with power law indices of Section B-C reported to be $\alpha_1\simeq 1.58_{-0.86}^{+0.82}$ and, respectively, $\alpha_2\simeq 5.6_{-2.6}^{+4.1}$.

Fig.\,\ref{fig:plaw} quantifies the conversion of observed-to-true power-law indices in
GWTC-2. For the tail of interest, it shows
\begin{equation}
\frac{\alpha_{B,true}}{\alpha_{B,obs}} \simeq \frac{4.98}{4.26}\simeq 1.17,
\label{EQN_alphaB}
\end{equation}
ignoring systematic uncertainties in this conversion process.
Revisiting uncertainty by scatter given the relatively modest number of events in the tail of the
present survey, Fig.\,\ref{fig:hist} repeats the estimate of $\alpha_B=\alpha_{B,obs}$
using four bins in a loglog plot of event count versus $\mu$. By (\ref{EQN_alphaB}), we 
infer 
\begin{eqnarray}
\alpha_B = 4.08\pm 0.73,~\alpha_{B,true}=4.77\pm 0.73.
\label{alphaB}
\end{eqnarray}
Our estimate is hereby consistent with but slightly less than 
$\alpha_2=\alpha_{B,true}$ of \cite{2021ApJ...913L...7A}. In particular,
$\alpha_{B,true}$ satisfies the astrophysical bound discussed in the following section.

\section{Astrophysical bounds on power-laws}

The observed steep index in Eq.~(\ref{EQN_alphaB}) satisfies some priors derived from aforementioned Salpeter IMF
as follows. From Eq.~(\ref{eq:imf}), the progenitor binary mass-function $M_1^\prime$ and $M_2^\prime$ of the BBH can be expressed as a function of mean $\mu^\prime$ and normalized mass difference $x=\nu^\prime/\mu^\prime$, $\nu^\prime=(M_1^\prime-M_2^\prime)/2$. (Equivalently, consider $\mu^\prime$ and $Q=2x^2/(1-x^2)$, $x=\nu^\prime/\mu^\prime$ in symmetrized mass-ratio $Q=(q+1/q)/2-1, q=M_2^\prime/M_1^\prime$, $\nu^\prime=\delta M^\prime$.) The two-parameter binary mass-function $\psi_B(\mu^\prime,x)$ has two natural limits derived from essentially {\em correlated} or {\em uncorrelated} stellar masses.

On the first, we note that nearby Galactic open stellar clusters show a large fraction of O-type stars born as binaries \citep{sana2012binary}. Radiation-hydrodynamic simulation suggests fragmentation of rotating gas disk 
by gravitational instabilities leads to the formation of massive binary stars with similar masses \citep{krumholz2009formation,sana2012binary,moe2017mind,2022arXiv220101905L}.
On the other hand, such mass-correlation is expected to weaken as binary separation becomes large. Accordingly, therefore, the binary stellar mass-function is expected to be bounded by either of the two limits
{
\begin{align}
\psi_B \propto \Bigg\{
        \begin{array}{ll}
        (\mu^\prime)^{-\alpha_S}, \\\\ 
        \left[ (\mu^\prime)^2-(\nu^\prime)^2\right]^{-\alpha_S} 
        \end{array}
        \label{eq:psi}
\end{align}
for the correlated and, respectively, uncorrelated case with $M_{1,2}^\prime=\mu^\prime \pm \nu^\prime$. 
}

{The IMF in $\mu^\prime$ of uncorrelated binary masses assumes a power-law index distinctly steeper than $\alpha_S$ in the correlated case upon marginalization over $x=\nu^\prime/\mu^\prime =(1-q)/(1+q)$ given fluctuations in mass ratio $q$.}
{The result of integrating out $\nu^\prime$ depends on the expected astrophysical range of $q$. Following a change of variables $dM_1^\prime dM_2^\prime =2d\mu^\prime d\nu^\prime  = 2 \mu^\prime d\mu^\prime dx$, $\nu^\prime=\mu^\prime x$, $M_{1,2}^\prime=\mu^\prime\left(1\pm x\right)$, an index 
$2\alpha_S-1$ obtains after integration over $0\le x \le 1$ covering all of $0\le q \le 1$. However, such includes extreme mass ratios $q=0$, i.e., $M_2=0$, $\nu=\mu$ ($x=1$). This limit is ruled out by a lower bound on the mass of black hole progenitor stars. Excluding this suggests, alternatively, the approximation $(\mu^\prime)^2-(\nu^\prime)^2\simeq (\mu^\prime)^2$. Integration over a finite strip in $\nu^\prime$, uncorrelated to $\mu^\prime$, produces an index $2\alpha_S$.}

{According to the above,} we expect $\psi_B \propto \left(\mu^\prime\right)^{-\alpha_{B,true}^\prime}$ with 
\begin{equation}
\alpha_S \lesssim \alpha_{B,true}^\prime \lesssim 2\alpha_S.
\label{EQN_AS}
\end{equation}

The index Eq.~(\ref{EQN_alphaB}) is significantly steeper than aforementioned Salpeter IMF with $\alpha_{S}$ = 2.35 of their progenitor stellar mass-function. In fact, Eq.~(\ref{EQN_alphaB}) shows an index essentially equal to {\em twice the Salpeter value of a binary IMF with uncorrelated masses}.

Some of the steepening in Eq.~(\ref{EQN_alphaB}) might alternatively be attributed to cosmological time-dilation. 

\section{Steepening by cosmic time-dilation}\label{sec:cosm}

For a given cosmological background evolution, redshifts $z^\prime$ of BBH progenitors can be traced back by times of coalescence (\citealt{peters1964gravitational,celoria2018lecture,mapelli2018astrophysics}), $t_{c} = (5/256)c^{-5}G^{-3}a^4 \left(M_1 M_2\,(M_1 + M_2)\right)^{-1}$, where $c$ is the speed of light, $G$ is the gravitational constant, $a$ is the initial separation, and $M_1$ and $M_2$ are mass of the primary BH and that of secondary BH. That is,
\begin{equation}
{t_{c}} \approx 0.1576\,\left[  \frac{ a } { 0.2 AU }  \right]^{4}\,\left[ \frac{\mu} {50 M_{\odot}} \right]^{-3} {t_{H}},
\label{eq:tcoal}
\end{equation}
where $\mu$ is the mean mass, and $t_{H}=13.7\,$Gyr is the Hubble time.

The PDF in the source frame, $P^\prime\left(\mu,z^\prime\right)$ 
= $P\left(\mu,z\right)$ $(\partial z/\partial z^\prime)\left(1+z^\prime\right)$,
derives from invariance of count $N$ and $\mu$, where $dN = {\psi}d{\mu}dT$ for an IMF $\psi$ and time interval $T$, whereby $P({\mu},z)d{\mu}dzdt = P'({\mu},z')d{\mu}dz'dt'$. Consequently, the transformation $(z,\mu)\rightarrow(z^\prime,\mu)$ with $z^\prime=z^\prime(z,\mu)$ gives
$P({\mu},z^\prime) (\partial({\mu},z)/\partial({\mu},z^\prime)) d{\mu}dz^\prime=P^\prime({\mu},z^\prime)Jd{\mu}dz^\prime$ with Jacobian $J = \partial z/\partial z'$, where $z$ and $z^\prime$ are related on given cosmological background evolution. Hence, $P^\prime({\mu},z^\prime) = P({\mu},z)(\partial z/\partial z^\prime)(dt/dt^\prime)$, i.e.:
\begin{equation}
P^\prime({\mu},z^\prime) = P({\mu},z) J_z(z) 
\label{EQN_P}
\end{equation}
as asserted, where $dt/dt^\prime=1+z^\prime$, with the cosmic steepening factor 
\begin{equation}
J_z = \frac{\partial z} {\partial z^\prime}\left(1+z^\prime\right).
\label{EQN_P2}
\end{equation}

For a concrete illustration, we consider the following Ansatz for progenitor redshift $z'$:
\begin{equation}
a = f \left(\frac{\mu}{\mu_0}\right)^{\gamma}
\label{eq:m_a}
\end{equation}
in the two parameters $f> 0$ and $\gamma>0$. Here,
$\mu_0=31.4\,M_{\odot}$ is the minimum of the tail
of the BBH mass-function (Fig.~\ref{fig:plaw}).
For an initial separation $a$, the merging time $t_{c}$ satisfies
\begin{equation}
t_c \propto \frac{a^4}{\mu^3}\,\,{\displaystyle \propto }\,{\mu}^{4\gamma-3}
\label{eq:f_gamma}
\end{equation}
with positive ($\gamma>0.75$), negative ($\gamma<0.75$) or neutral ($\gamma=0.75$)
correlation between $t_c$ and $\mu$. For the PDF of the observed mergers, we consider
\begin{equation}
    P(\mu)\, {\displaystyle \propto }\, {\mu}^{-\alpha_{B}},
    \label{eq:pdf}
\end{equation}
where $\alpha_{B}$ is the observed value Eq.~(\ref{EQN_alphaB}), about the mean
redshift $\bar{z}=0.51$ in light of the modest standard deviation $\sigma_z=0.2$. Based on Eq.~(\ref{EQN_P}) and Eq.~(\ref{eq:f_gamma}), we numerically solve for $z^\prime$ 
of the progenitor given a merger event at $z$ on a three-flat $\Lambda$CDM background.

We highlight three model relationships between the initial separation $a$ and coalescence time $t_c$ with mass parameterized by $\gamma$ color-coded with blue, orange, and green, obtained with Python package {\sc cosmology} with Hubble parameter $H_0 = 67.8$\,km\,s$^{-1}$\,Mpc$^{-1}$ and matter density $\Omega_{m} = 0.307$. 

\begin{figure}
\centering
\includegraphics[width=0.95\columnwidth]{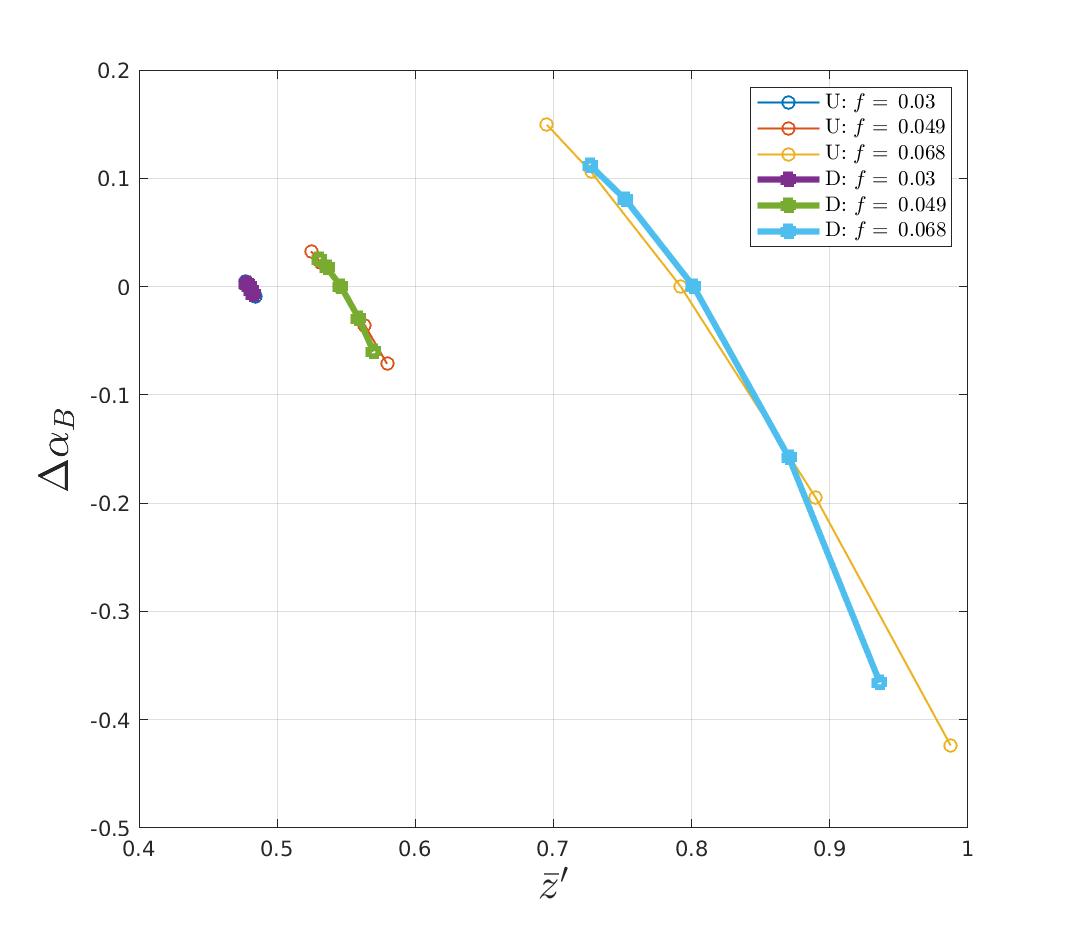}
\caption{Steepening in the power-law index $\Delta\alpha_B$ reaches about 8\% of (\ref{EQN_alphaB}) in progenitor redshift $z^\prime$ by cosmic time-dilation factor $J_z$ in the ansatz (\ref{eq:m_a}) over the index range $0.5\le \gamma\le1$ on a three-flat $\Lambda$CDM background, shown are for $f=0.03, 0.049, 0.068$. 
Results for a fiducial uniform distribution ($U$) and the BBH data in the tail of the GWTC-2 catalogue ($D$) are rather similar. 
}
\label{figR1}
\end{figure}

For illustrative purposes, Fig.~\ref{figR1} shows the change $\Delta \alpha_B=\alpha_B-\alpha_B^\prime$ by $J_z(z)$ (\ref{EQN_P2}) in (\ref{EQN_P}) for the data $(\mu,z)$ at hand, including, for illustrative purposes, the same for a uniform distribution in mass, $\mu_0\le \mu \le \mu_1$ ($\mu_0=31.4M_\odot$, $\mu_1=59.5M_\odot$) and redshift, $z_1\le z\le z_2$ ($z_1=0.42$, $z_2=0.52$), preserving similar mean values to those of the tail of the BBH distribution 
(Fig.~\ref{fig:plaw}).
 
\begin{figure}
\vskip0.2in

\includegraphics[width=0.94\columnwidth]{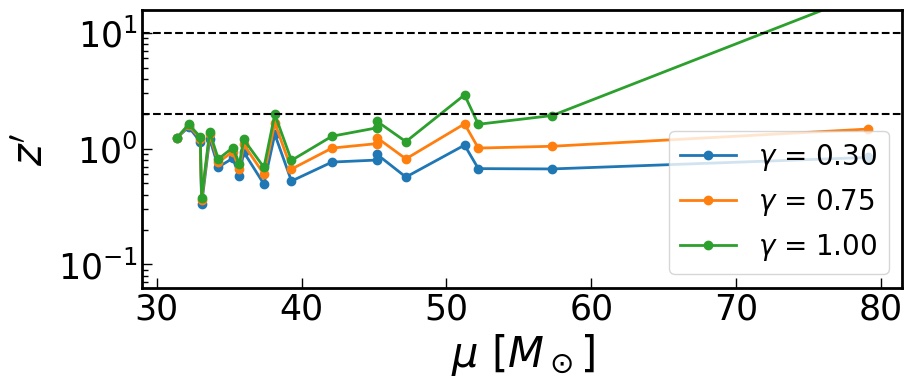}
\caption{
Progenitor redshifts $z^\prime$ for $\gamma=0.30$ (blue), $\gamma=0.75$ (orange) and $\gamma=1.00$ (green) in Eq.~(\ref{eq:m_a}) on a three-flat $\Lambda$CDM cosmological background. Horizontal dashed lines highlight $z^\prime = 2$ and 10. The associated mean of $z^\prime$ versus model parameter $\gamma$ increases above the mean redshift $z=0.51$ in the tail of the BBHs of GWTC-2 for $\gamma>0.75$.
$\alpha_B^\prime$ of their PDFs is summarized in Table~\ref{tab:three_cases},
showing steepening when $\mu-t_c$ satisfies a positive correlation ($\gamma > 0.75$).
}
\label{figR2}
\end{figure}

Figs.~\ref{figR2} shows the results for (\ref{EQN_P}). 
Long merger times push the origin of high-mass binaries to high $z^\prime$. In treating the (tail of the) BBH of GWTC-2 as a uniform population, the condition $z^\prime<\infty$ sets a bound $f\lesssim 0.15$. Shown is $z^\prime$ versus $\mu$ 
for selected values $\gamma$ = 0.30, 0.75, and 1.00 (Case 1-3), in blue, orange, and green, respectively.

Table~\ref{tab:three_cases} summarizes the steepening in the PDF$^\prime$ in the source frame compared to the observed PDF.
While subtle (hard to discern by eye), the steepening present in (\ref{EQN_P}) is consistent with the anticipated modest change due to $J_z$ shown in Fig.~\ref{figR1}.
Steepening appears for $\gamma>0.75$ (Case 3), otherwise absent for $\gamma\le 0.75$ (Case 1-2).

\begin{table*}
    \centering
    \begin{tabular}{ccccccccccccccc}
      \hline \hline
      \multirow{2}{*}  {\centering Case}  &  \multirow{2}{*}{Value} &  \multirow{2}{*}{$\mu\--t_{c}$ relation} & \multicolumn{3}{c}{$f = 0.150$} & \cline{4-6}   & & & $\alpha_B^\prime$ & $\alpha_{B,true}^\prime$ & $\bar{z^\prime}$ \\ 
       \hline 
        1 & $\gamma < 0.75$ & negatively correlated & $4.10_{-0.73}^{+0.74}$ & $4.80_{-0.73}^{+0.74}$   & 0.86 \\ 
       2 & $\gamma = 0.75$ & constant                     & $4.10_{-0.74}^{+0.73}$  & $4.80_{-0.74}^{+0.73}$   & 1.07 \\
      3 & $\gamma > 0.75$ & positively correlated   & $4.28_{-0.74}^{+0.73}$   & $5.00_{-0.74}^{+0.73}$   & 2.38\\
       \hline
\end{tabular}
 \caption{Estimated power-law index $\alpha_B^\prime$ in the source frame for a model relation of merging time scale $t_c$ and mass parameter $\mu$ of the binaries for three parameter values of $\gamma$ and two $f$ values. The case $\gamma > 0.75$ demonstrates steepening in $\alpha_B^\prime$ of the progenitor systems relative to the index $\alpha_B$ of the mass-function at coalescence. }
\label{tab:three_cases}
\end{table*}

Steepening by cosmic time-dilation is noticeable when progenitor and merger redshifts differ by order unity $\left(z^\prime-z\gtrsim 1\right)$. Such may push $\alpha_B^\prime$ across the upper bound $2\alpha_S$ in Eq.~(\ref{EQN_AS}) in the binary IMF of progenitor stellar systems. Avoiding this limits the origin of the BBH progenitors to relatively low-$z$ late-time cosmology, 
effectively posterior to the peak in the cosmic star formation rate (\citealt{hopkins2006normalization,2014ARAA..52..415M}).

\section{Discussion and conclusions}\label{sec:summary}

The BBH mass-function of LIGO BBH mergers is identified with the Salpeter IMF of their stellar progenitors in the power-law tail $\mu \geq 31.4 \,M_\odot$ (Fig.~\ref{fig:plaw}), effectively parameterized by mean mass $\mu$ due to the implied tight correlation with chirp mass Eq. (\ref{EQN_kappa}). 
The mass scale Eq.~(\ref{EQN_geo1}) expected from the Salpter IMF, subject to preserving binary association in the final phase of binary evolution, is consistent with the observed masses in the tail shown in Figs.~\ref{fig:plaw}-\ref{fig:hist}.

Following a detailed consideration of GWTC-2 data, our main findings indicate
\begin{enumerate}
\item 
A tight correlation Eq.\,(\ref{EQN_r}) between primary and secondary masses, consistent with the paucity in intermediate mass X-ray binaries; 
\item 
A broken power-law mass-function with a tail beyond $\mu\gtrsim 31.4M_\odot$. The power-law index $\alpha_{B,true}\simeq 4.77\pm 0.73$ is consistent with uncorrelated stellar progenitor masses at birth Eq.~(\ref{eq:psi}) by approaching the limit $2\alpha_S=4.7$ defined by the Salpeter index $\alpha_S$;
\item
A power-law index of $\mu$ is subject to steepening due to cosmological time-dilation (Fig.~\ref{figR1}), e.g., when mass and orbital separation are positively correlated. 
The condition 
Eq.\,(\ref{EQN_AS}) hereby bounds the mean of progenitor redshift $\bar{z}^\prime$.
\item 
Subject Eq.\,\ref{EQN_AS}, Table \ref{tab:three_cases} suggests that a progenitor origin 
at $z^\prime\gg1$ is excluded, assuming the tail of the BBH population to derive from a uniform population. Progenitors hereby appears to be in the relatively recent epoch of cosmic star formation, about or posterior to the peak in the star formation rate. 
\end{enumerate}

Conceivably, the last finding can be made more rigorous with BBH surveys from upcoming O4-O5 observations. Such bounds hold promise to distinguish between an origin related to the peak in the cosmic star formation rate and an association with Pop III stars \citep[e.g.][]{yajima2015can,kulkarni2014chemical,kinugawa2014possible}, previously considered for their relatively high mass of several tens of $M_\odot$ \citep{kinugawa2014possible,hosokawa2011protostellar,kinugawa2016the,kinugawa2021gravitational}.

The tight correlation between primary and secondary BH masses and the uncorrelated binary progenitor masses in our findings is perhaps paradoxical. However, the pathway to black hole binaries from stellar progenitor systems is a complex process of binary stellar evolution including (uncertain) mass-losses in stellar winds \citep[e.g.][]{kriticka2014mass,chen2015parsec,belczynski2012missing,elbert2018counting}, mass-transfer potentially equalizing masses \citep[e.g.][]{kinugawa2014possible}, terminating in two core-supernovae. 
These processes are subject to a stringent selection criterion of preserving binary association, i.e., a 50\% mass-loss limit (in the idealized case of circular binary motion). 

Equalizing masses in the progenitor systems offers some hints at their stellar evolution, perhaps with further contributions to steepening in the power-law index from wind mass-loss \citep[cf.][]{vink2001mass}; \citealt{vink2011wind}; \citealt{chen2015parsec}), e.g., in binary association around cold red giants \citep{decin2020stellar}. While a detailed study of these complex radiation-hydrodynamical processes is outside the scope of this work, these processes and their down-selection effects in binary black hole formation are probably instrumental to understanding the detailed nature of the progenitor systems, here identified with ab initio uncorrelated progenitor stellar binary masses in view of $\alpha_B\simeq 2\alpha_S$ in Eq.~(\ref{EQN_r}).

\mbox{}\\
{\bf Data Availability.} The data underlying this article were accessed from LIGO. 

\mbox{}\\
{\bf Acknowledgments.} The authors thank the anonymous reviewer for a detailed reading and constructive comments. This research is supported, in part, by NRF of Korea Nos. 2015R1D1A1A01059793, 2016R1A5A1013277 and 2018044640. Shinna Kim and Shin-Jeong Kim acknowledge a support from the National Research Foundation of Korea (NRF) grant funded by the Korea government (Ministry of Science and ICT: MSIT) (No. NRF-2022R1A2C1008706).

\bibliographystyle{mnras}
\bibliography{msR2}

\end{document}